\documentclass[twocolumn,showpacs,preprintnumbers,amssymb]{revtex4}
\usepackage{pdfpages}
\usepackage{graphics}
\usepackage{epsfig} 
\usepackage{hyperref}
\usepackage{graphics}
\usepackage{color}      % use if color is used in text
\usepackage{graphicx}
\usepackage{graphicx}% Include figure files
\usepackage{dcolumn}% Align table columns on decimal point
\begin{document}

 \title{Noise-Activated Dopant Dynamics in Two-Dimensional Thermal Landscapes with Localized Cold Spots}

\author{Mesfin Asfaw Taye}
\affiliation{West Los Angeles College, Science Division,\\ 9000 Overland Ave, Culver City, CA 90230, USA}
\email{tayem@wlac.edu}

\author{Yoseph Abebe}
\affiliation{Department of Physics, Debre Markos University, Debre Markos, Ethiopia}

\author{Tibebe Birhanu}
\author{Lemi Demeyu}
\author{Mulugeta Bekele}
\affiliation{Department of Physics, Addis Ababa University, P.O. Box 1176, Addis Ababa, Ethiopia}

\begin{abstract}
Controlling dopant transport with high spatial precision is crucial for improving the semiconductor functionality, reliability, and scalability. Although prior models of noise-assisted diffusion have been largely confined to idealized one-dimensional settings, we present a physically realistic two-dimensional theoretical framework that integrates anisotropic quartic confinement with localized thermal cold spots to direct impurity dynamics. Using a generalized Fokker–Planck formalism, we show that the geometry of the thermal landscape, particularly the width and arrangement of cold spots, governs a noise-induced transition between monostable and bistable effective potentials. This enables tunable noise-activated hopping and supports conditions favorable for stochastic resonance (SR) if weak periodic driving is applied. Quantitative predictions are made for how impurity localization and effective barrier heights depend on the cold-spot width $\sigma$ and trap depth $\Phi$, offering experimentally testable signatures. We propose an experimental realization using optothermal techniques, such as dual-beam optical tweezers and laser cooling, which can sculpt reconfigurable thermal profiles with sub-micron resolution. This model establishes a versatile pathway for programmable impurity manipulation and noise-sensitive control in semiconductor structures, bridging theoretical predictions with feasible experimental detection via photoluminescence mapping or lock-in signal amplification.\end{abstract}

\pacs{Valid PACS appear here}% PACS, the Physics and Astronomy
\maketitle

%\section{Introduction}

\textit{Introduction.\textemdash}  
Doping plays  an important  role in semiconductor technology  since it   fundamentally  governs  the carrier concentration, electrical conductivity, and overall device performance in transistors, sensors, and photovoltaic systems~\cite{37,38,39,40,41}. Attaining  precise spatial and temporal control over dopant distributions is essential for tailoring the electronic, optical, and transport properties. However, this  remains challenging  because of competing mechanisms among  vacancy hopping, interstitial migration, and drift driven by electric and thermal fields~\cite{15,16,17}. Among these, thermally activated diffusion is dominant because  temperature gradients play a critical role in the redistribution of impurities over time~\cite{16,20}.  

While noise is commonly regarded as a disruptive factor in signal transmission, it can, under the right conditions, enhance transport via stochastic resonance (SR)~\cite{1,2,3,4,5}, which was originally proposed to explain Earth’s glacial cycles~\cite{7}. In SR, noise synchronizes with weak periodic inputs to amplify the system response~\cite{6}, an effect observed in confined geometries~\cite{8}, polymer dynamics~\cite{9,10,11,12}, porous membranes~\cite{13}, and bistable two-state kinetics~\cite{1,14}. In semiconductors, thermal gradients coupled with noise offer new strategies for the high-resolution control of dopants. Recent studies have shown that localized thermal features (hot or cold spots) can induce bistability and enable the selective redistribution of dopants~\cite{18,19,20,21} which  facilitates noise-assisted signal processing in semiconductor layers.

Previous models largely focus on one-dimensional systems with quartic confinement and spatially varying temperature fields~\cite{18,19,21,25}, analyzed through generalized Fokker--Planck or Langevin formalisms and capturing thermally activated hopping via modified Arrhenius kinetics~\cite{22,23,24,25,26}. These models predict noise-induced transitions between monostable and bistable impurity configurations, which is a prerequisite for SR under weak periodic forcing~\cite{27,28,29}, and quantify signal amplification via spectral gain and signal-to-noise ratio~\cite{1,14,30}.

Inspired by advances in nanoscale gating and thermal control \cite{30,31}, we present a novel two-dimensional framework that captures impurity dynamics in realistic planar geometries by integrating anisotropic quartic confinement with spatially localized thermal cold spots. We demonstrate that the cold-spot width \(\sigma\) and entropy-modulated trap depth \(\Phi\) determine the crossover between the monostable and bistable effective potentials. These transitions govern noise-activated hopping, impurity localization, and barrier heights, offering experimentally accessible predictions. Our proposed configuration, implementable via optical tweezers~\cite{33,36}, laser cooling~\cite{35}, and spatial light modulation~\cite{25,33,34}, enables programmable impurity control and supports conditions that are favorable for SR. This model bridges the idealized stochastic transport theory with experimentally realizable two-dimensional systems, contributing a robust mechanism for precision doping and noise-enhanced functionality in semiconductor platforms.

{\it The model and effective potential.\textemdash}
We investigate the stochastic motion of dilute impurities in a two-dimensional confined semiconductor exposed to spatial temperature gradients and external trapping. This setting captures the essential features of experimentally accessible systems such as thin-film devices and optically manipulated colloids~\cite{33,34,35}.
The external confinement is modeled by a symmetric quartic potential,
\begin{equation}
V(x, y) = V_0 \left[ \left( \frac{x}{x_{\max}} \right)^4 + \left( \frac{y}{y_{\max}} \right)^4 \right],
\end{equation}
where \( V_0 \) defines the energy scale and \( x_{\max}, y_{\max} \) sets the spatial extent. This smooth potential shape emulates the soft-trapping profiles used in optical and electrostatic confinement platforms.

To model thermal inhomogeneity, we impose a temperature field with two symmetric cold regions centered at \( x = \pm x_0 \),
\begin{equation}
T(x, y) = \frac{T_0}{1 + \exp\left[ \frac{(x + x_0)^2 + y^2}{2\sigma^2} \right] + \exp\left[ \frac{(x - x_0)^2 + y^2}{2\sigma^2} \right]},
\end{equation}
where \( T_0 \) is the ambient temperature, \( \sigma \) controls the width of each cold spot, and \( x_0 \) sets the separation.
Note that this thermal profile can be  created  using laser-based thermoplasmonics, and in this case,  focused beams generate localized heating and thereby define effective cold zones via contrast. Alternative approaches include uniform illumination with masked cooling or patterned absorptive layers. Practical implementations may also  employ spatial light modulators, dual-beam optical tweezers, or tailored coatings on thermally conductive substrates~\cite{33,36}.

The dynamics of the impurity ensemble are governed by a generalized Fokker--Planck equation 
\begin{equation}
\frac{\partial P}{\partial t} = \nabla \cdot \left[ \frac{\nabla V(x, y)}{k_B T(x, y)} e^{-\Phi / k_B T(x, y)} P + \nabla \left( e^{-\Phi / k_B T(x, y)} P \right) \right],
\end{equation}
where \( P(x, y, t) \)  denotes  the probability density.  The parameter 
 \( \Phi \) represents the energy cost associated with impurity motion such as binding affinity or effective interaction with the medium. It also  modulates the response to local temperature variations.
At steady state, the probability distribution becomes
\begin{equation}
P(x, y) = C \cdot e^{-\Phi / k_B T(x, y)} \cdot \exp\left( - \int \frac{\nabla V(x', y')}{k_B T(x', y')} \cdot d\mathbf{r}' \right),
\end{equation}  and this 
illustrates the combined effects of energetic confinement and thermal modulation on impurity localization.
To provide physical insight, we define an emergent effective potential \( V_{\mathrm{eff}}(x, y) \) through the relation \( P(x, y) \propto e^{-V_{\mathrm{eff}}(x, y)/k_B T_0} \), yielding
\begin{equation}
V_{\mathrm{eff}}(x, y) = \frac{\Phi T_0}{T(x, y)} + k_B T_0 \int_{(x_0, y_0)}^{(x, y)} \frac{\nabla V(x', y')}{T(x', y')} \cdot d\mathbf{r}'.
\end{equation}
This effective potential captures both entropy-suppressing effects near cold regions and spatial confinement due to the external potential. For narrow cold spots (\( \sigma \ll x_0 \)), the landscape exhibits a double-well structure,  which in turn promotes  bistability and noise-driven hopping. As \( \sigma \) increases or \( \Phi \) decreases, the barrier flattens, and the system transitions to monostability. Experimental platforms such as dual-beam optical tweezers and thermoplasmonic heating~\cite{33,35,36} offer precise in situ control over parameters  such as \( \sigma \), \( \Phi \), and \( V_0 \).  As a result,  this enables  direct implementation of noise-assisted transport and thermal regulation of impurities.

\begin{figure}[htbp]
\centering
\begin{tabular}{cc}
\includegraphics[width=0.25\textwidth]{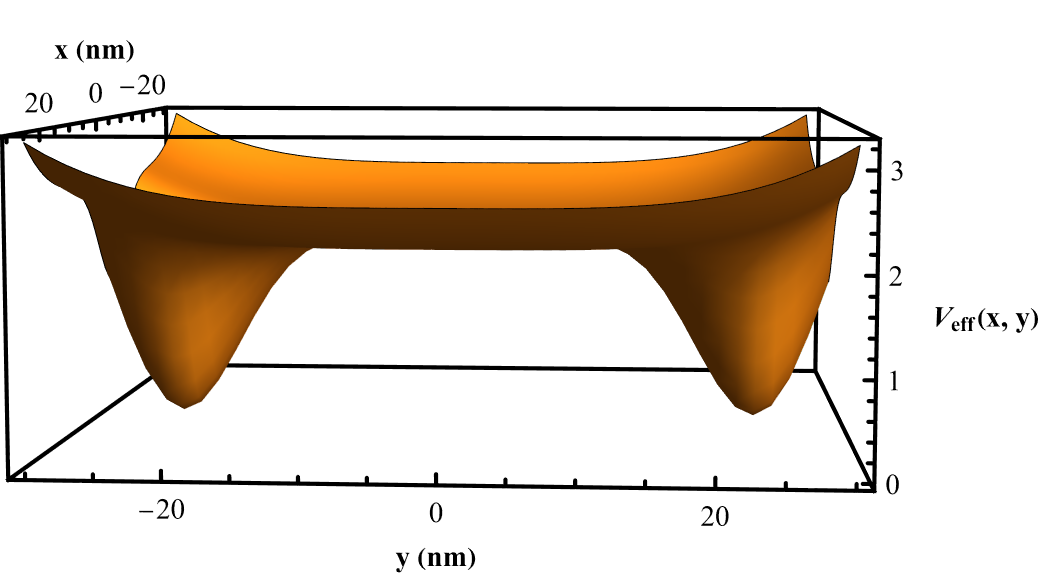} &
\includegraphics[width=0.25\textwidth]{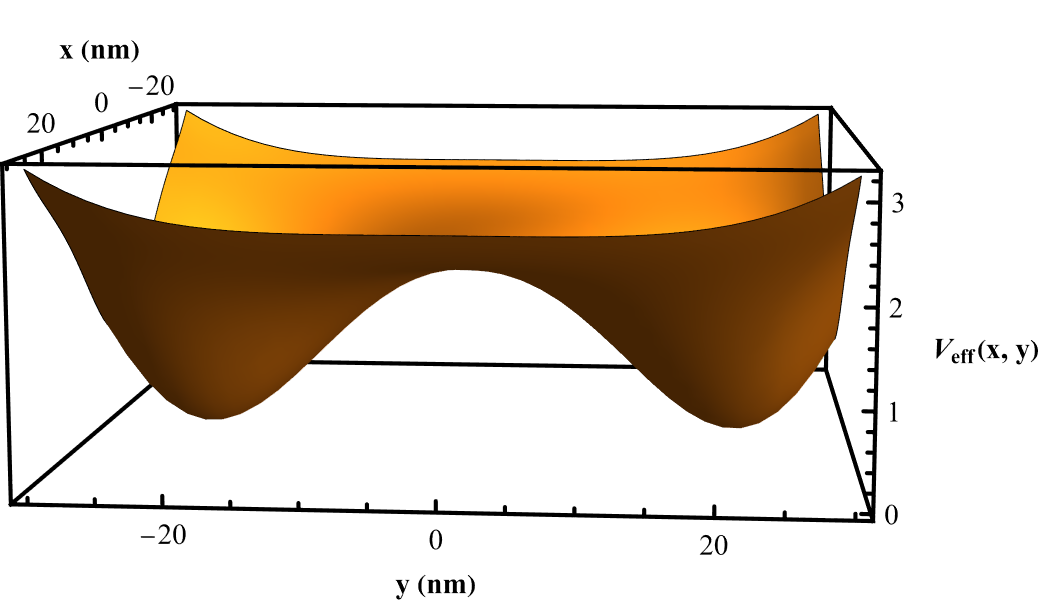} \\
(a) $\sigma = 5\,\mathrm{nm}$ & (b) $\sigma = 10\,\mathrm{nm}$
\end{tabular}
\caption{
(Color online) Contour plots of effective potential $V_{\mathrm{eff}}(x, y)$ for two different cold-spot widths under a fixed background temperature $T_0 = 1.0$, trap depth $\Phi = 10.5\, \mathrm{eV}$, and confinement scale $x_{\max} = y_{\max} = 25\,\mathrm{nm}$. In panel (a), a narrow cold spot ($\sigma = 5\,\mathrm{nm}$) creates a sharply defined bistable potential, while in panel (b), the broader spot ($\sigma = 10\,\mathrm{nm}$) reduces s the potential landscape, reducing the bistability.}
\label{fig:veff_sigma_comparison}
\end{figure}

\begin{figure}[htbp]
\centering
\begin{tabular}{cc}
\includegraphics[width=0.25\textwidth]{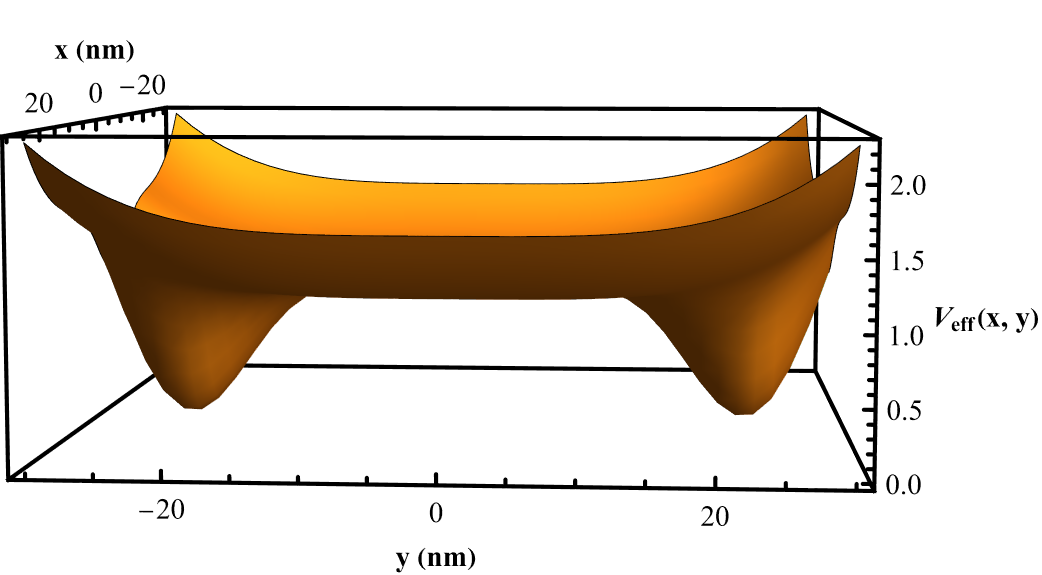} &
\includegraphics[width=0.25\textwidth]{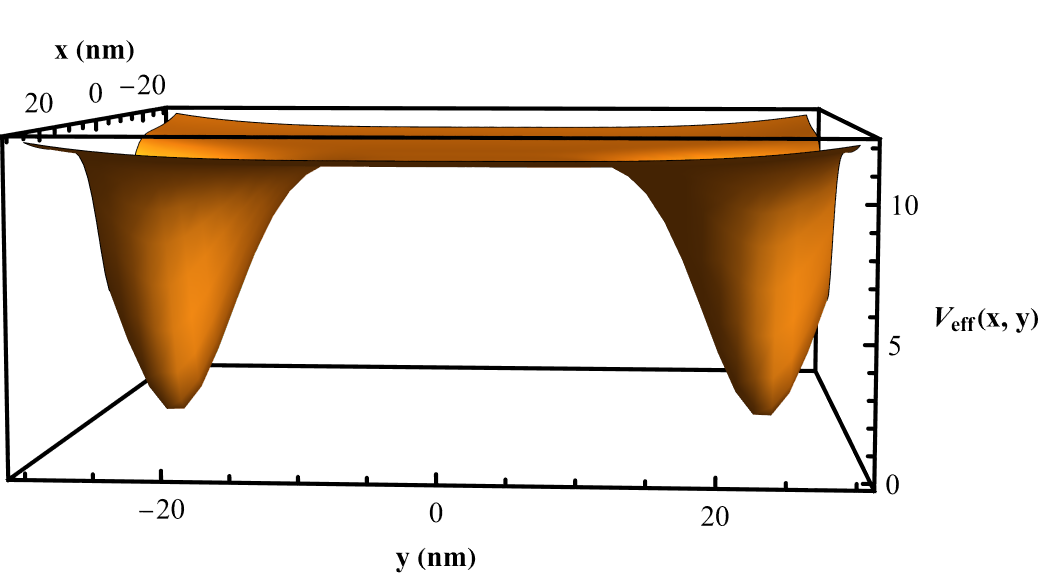} \\
(a) $\Phi = 0.05\,\mathrm{eV}$ & (b) $\Phi = 10.5\,\mathrm{eV}$
\end{tabular}
\caption{
(Color online) Contour plots of the effective potential $V_{\mathrm{eff}}(x, y)$ illustrating the effect of trap depth $\Phi$ at a fixed cold-spot width $\sigma = 5\, \mathrm{nm}$, background temperature $T_0 = 1.0$, and quartic confinement with $x_{\max} = y_{\max} = 25\, \mathrm{nm}$. Panel (a) shows a weakly modulated potential for shallow trapping ($\Phi = 0.05\,\mathrm{eV}$), whereas panel (b) demonstrates strong impurity localization and deeper bistability with $\Phi = 10.5\,\mathrm{eV}$. Increasing $\Phi$ amplifies the spatial contrast in $V_{\mathrm{eff}}$, enhancing confinement at the cold spots.}
\label{fig:veff_phi_comparison}
\end{figure}

\begin{figure}[htbp]
  \centering
  \includegraphics[width=0.5\textwidth]{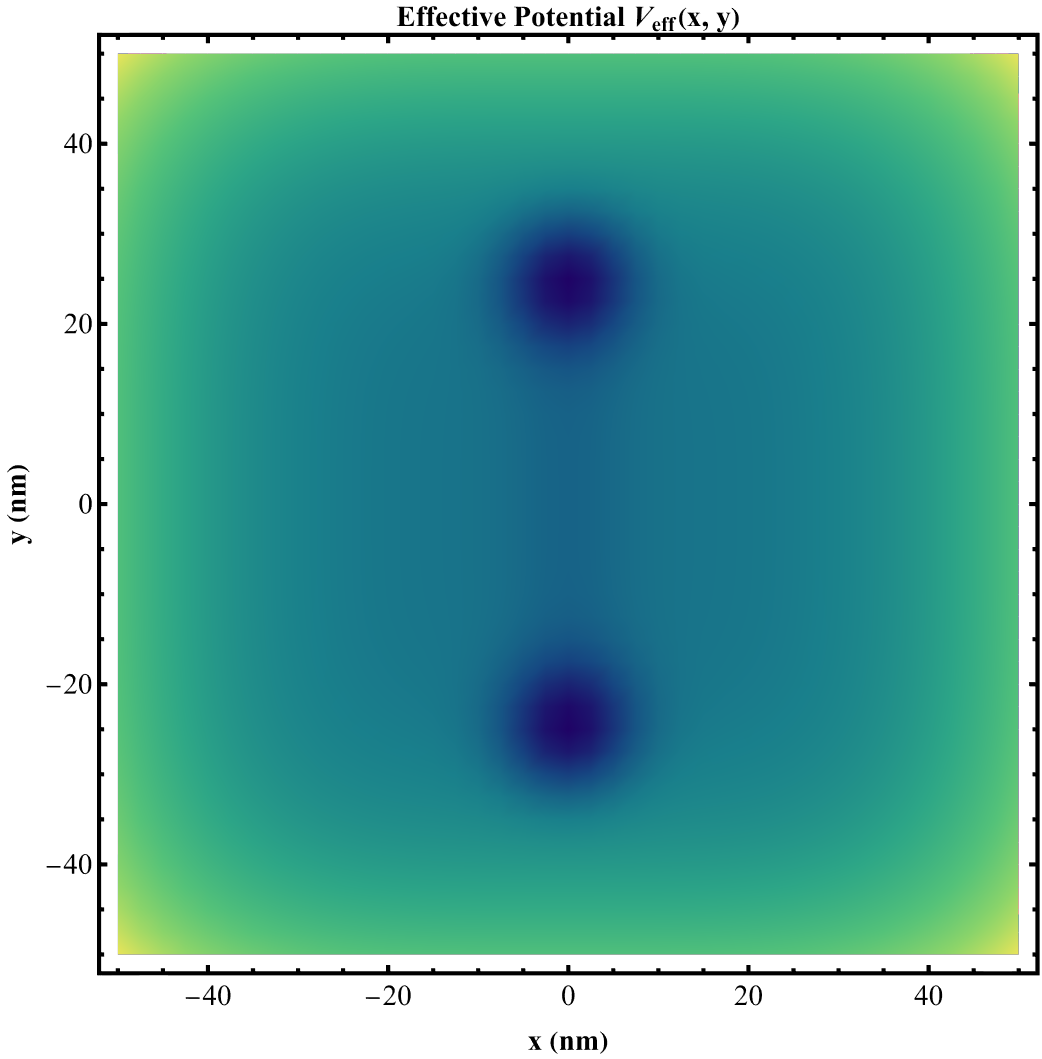}
  \caption{Contour plot of the corrected effective potential $V_{\mathrm{eff}}(x, y)$ for a semiconductor layer under thermal confinement. The potential includes both symmetric quartic confinement with $x_{\max} = y_{\max} = 25\, \mathrm{nm}$ and two localized cold spots centered at $x = \pm x_0$ with $x_0 = 25\, \mathrm{nm}$ and spatial width $\sigma = 5\, \mathrm{nm}$. The background temperature is uniform at $T_0 = 1.0$ (arb. units), and the thermal trap correction has a depth $\Phi = 2.5\, \mathrm{eV}$. The effective potential is computed using path-integrated gradients of both the confining potential and temperature landscape, yielding a bistable configuration characterized by two symmetric wells and a central depletion zone.}
  \label{fig:veff_contour_phi2p5}
\end{figure}

\begin{figure}[htbp]
  \centering
  \includegraphics[width=0.5\textwidth]{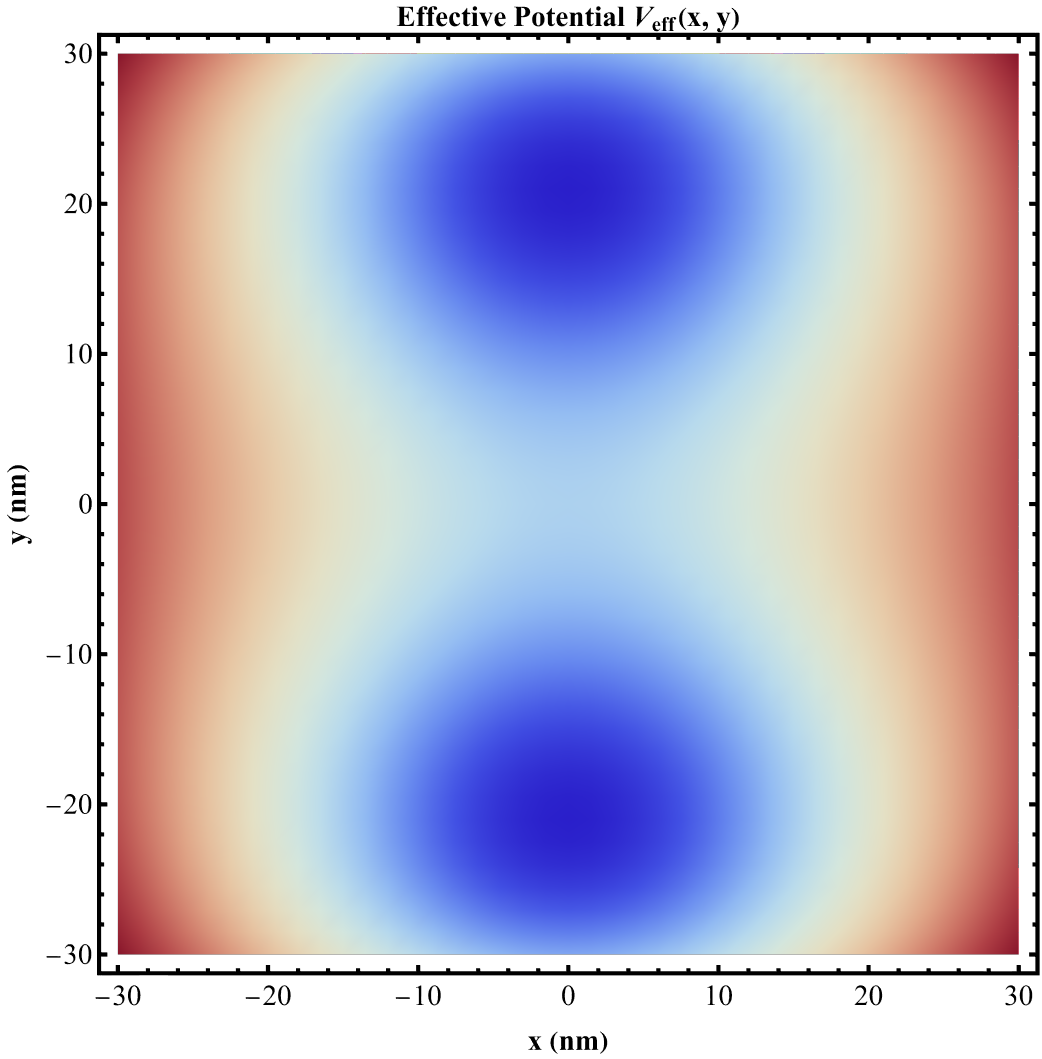}
  \caption{Contour plot of the corrected effective potential $V_{\mathrm{eff}}(x, y)$ for an impurity system in a semiconductor layer. The system experiences symmetric quartic confinement with $x_{\max} = y_{\max} = 25\, \mathrm{nm}$ and background temperature $T_0 = 1.0$ (arb. units). Two symmetric cold regions are applied at $x = \pm x_0$ with $x_0 = 25\, \mathrm{nm}$ and Gaussian width $\sigma = 15\, \mathrm{nm}$. The trap correction depth is set to $\Phi = 2.5\, \mathrm{eV}$. In contrast to the sharp localization observed at smaller $\sigma$, broader temperature wells result in a shallower effective potential landscape. The domain covers $x, y \in [-30, 30]\, \mathrm{nm}$.}
  \label{fig:veff_contour_sigma15}
\end{figure}

The spatial structure of the effective potential $V_{\mathrm{eff}}(x,y)$ strongly depends on the thermal profile parameters, particularly the width $\sigma$ of the cold spots and trap depth $\Phi$. These dependencies are systematically visualized in Figs. ~\ref{fig:veff_sigma_comparison}–\ref{fig:veff_contour_sigma15}, where the contour plots illustrate the resulting landscapes under different physical regimes.  Figure~\ref{fig:veff_sigma_comparison} depicts  the impact of the cold-spot width on the emergence of bistability. For a narrow cold spot with $\sigma = 5\,\mathrm{nm}$ [Fig. ~\ref{fig:veff_sigma_comparison}(a)], the effective potential develops two well-separated minima centered near $x = \pm x_0$ that form  a clearly bistable system with a high central barrier. This landscape favors noise-activated hopping and localization near cold zones. As the cold-spot width increases to $\sigma = 10\,\mathrm{nm}$ [Fig.~\ref{fig:veff_sigma_comparison}(b)], the thermal modulation becomes spatially smoother, leading to a significant reduction in the central barrier and flattening of the peripheral wells. The reduction in the barrier height between wells eliminates bistability and establishes a monostable configuration.

Figure~\ref{fig:veff_phi_comparison} shows the influence of the entropy-induced trap depth parameter $\Phi$ at a fixed $\sigma = 5\,\mathrm{nm}$. For weak modulation ($\Phi = 0.05\,\mathrm{eV}$), panel (a) shows a nearly uniform potential with shallow minima, indicating poor localization and enhanced thermal delocalization. In contrast, increasing $\Phi$ to $10.5\,\mathrm{eV}$ in panel (b) leads to deep potential wells at the cold spot locations, sharpening impurity confinement, and increasing the energy barrier for escape.

To further illustrate the combined effect of the cold-spot geometry and confinement, Figs. ~\ref{fig:veff_contour_phi2p5} and~\ref{fig:veff_contour_sigma15} present the full-domain contour plots of $V_{\mathrm{eff}}(x,y)$ computed via the numerical path integration of the temperature and potential gradients. In Fig.~\ref{fig:veff_contour_phi2p5}, we set $\Phi = 2.5\,\mathrm{eV}$ and $\sigma = 5\,\mathrm{nm}$ to capture the bistable configuration, where impurities experience strong localization near the two off-center minima. The central region acts as a thermal depletion zone, which is consistent with the noise-suppressed escape trajectories. By contrast, in Fig.~\ref{fig:veff_contour_sigma15}, we retain $\Phi = 2.5\,\mathrm{eV}$ but increase the cold-spot width to $\sigma = 15\,\mathrm{nm}$. This broadening eliminates the double-well structure and yields a smoother, nearly monostable landscape.

These results collectively demonstrate that the structure of $V_{\mathrm{eff}}(x,y)$, and hence the impurity localization behavior, can be finely controlled by tuning $\sigma$ and $\Phi$. The figures validate this control numerically and provide a visual basis for interpreting the crossover from bistable to monostable behavior. This control mechanism underlies the potential for noise-tunable impurity manipulation in optothermally patterned devices.

To gain a deeper insight into the confinement mechanism, we analyze the effective activation barrier that separates the cold spot minimum at \( x = x_0 \) from the central saddle at \( x = 0 \) along the symmetry line \( y = 0 \). This barrier characterizes the energetic cost of thermally activated hopping between wells and is approximately given by
\begin{equation}
\Delta V_{\mathrm{eff}} \approx \frac{\Phi \, \epsilon}{T_0} \left( 2 e^{-x_0^2 / 2\sigma^2} - 1 \right) + \frac{V_0 k_B x_0^4}{x_{\max}^4},
\label{eq:scaling_law}
\end{equation}
where the first term represents the entropy-induced contribution due to localized cooling, with \( \epsilon \) denoting the fractional temperature contrast and \( \sigma \) the spatial width of the cold spot. The second term arises from quartic mechanical confinement. The scaling in Eq. (6) reveals how sharper thermal gradients (smaller \( \sigma \)) and deeper cold regions (larger \( \epsilon \)) enhance localization by increasing the entropic barrier, whereas the quartic term grows algebraically with \( x_0^4 \), reinforcing spatial trapping. These contributions jointly govern whether the system supports bistability or collapses into a monostable configuration. Experimentally, such regimes can be realized using symmetric focused laser beams or optothermal fields to generate Gaussian-shaped cold spots at \( x = \pm x_0 \). The beam waist controls \( \sigma \), and the beam intensity regulates \( \epsilon \) through photothermal coupling. The full derivation of Eq. (6), along with detailed approximations and integration methods, are provided in Appendix~A.

To further interpret the effective potential across regimes, we consider several limiting cases. In the entropic or trap-dominated limit, where mechanical confinement is negligible, the potential reduces to \( V_{\mathrm{eff}}(x, y) \approx \Phi T_0 / T(x, y) \)  which captures  the localization governed purely by temperature gradients. 

Our proposed model, which integrates anisotropic quartic confinement with spatially localized thermal cold spots, offers a powerful framework for controlling dopant transport in semiconductor layers. Spatially, cold spots, realized via focused laser beams or thermoplasmonic techniques, generate localized temperature minima that act as tunable traps. By adjusting the lateral positions and widths (\(\sigma\)) of these cold regions, one can create customized potential well arrays that precisely shape impurity distributions. This  also enables targeted confinement  that suppresses diffusion into sensitive regions and supports the formation of spatially heterogeneous doping profiles. Moreover, dynamic control of the thermal landscape using holographic optical tweezers or digital micromirror devices allows real-time steering of dopants along programmable trajectories.
Temporally, the depth of the confinement potential (\(\Phi\)) regulates the activation energy required for dopant transitions between localized wells, thereby controlling the residence times and effective mobility. By tuning \(\Phi\), one can selectively enhance or suppress transport across specific regions  and this  enables  programmable temporal control over doping kinetics.

{\it Generalized Temperature Field and Effective Potential Landscape. \textemdash} We consider a two-dimensional system of non-interacting impurity particles diffusing under the combined influence of an external confining potential and a spatially modulated temperature field. To introduce tunable thermal asymmetries, the temperature profile is generalized to include multiple localized cold regions:
\begin{equation}
T(x, y) = \frac{T_0}{1 + \sum_i \exp\left[ -\frac{(x - x_i)^2 + (y - y_i)^2}{2\sigma^2} \right]},
\end{equation}
where \( T_0 \) is the far-field temperature, \( (x_i, y_i) \) are the cold spot centers, and \( \sigma \) sets the spatial extent.

Such structured thermal landscapes can be engineered experimentally using spatial light modulators (SLMs), holographic optical tweezers, or thermoplasmonic substrates. These methods allow precise laser shaping to locally alter the kinetic temperature of the medium and form symmetric or disordered cold zones. This flexibility enables the exploration of impurity localization in programmable or randomly distributed thermal environments.

As described in Eqs.~(4)–(5), the steady-state distribution \( P_{\mathrm{ss}}(x, y) \) depends on an effective potential \( V_{\mathrm{eff}}(x, y) \) that combines both the sensitivity to local temperature and the mechanical force landscape. Experimentally, by adjusting the number, width, or arrangement of cold spots, one can directly manipulate the minima of \( V_{\mathrm{eff}} \). This, in turn,  effectively shapes the impurity density profile and enables control over thermally guided transport and confinement.

\begin{figure}[ht]
\centering
\includegraphics[width=0.45\textwidth]{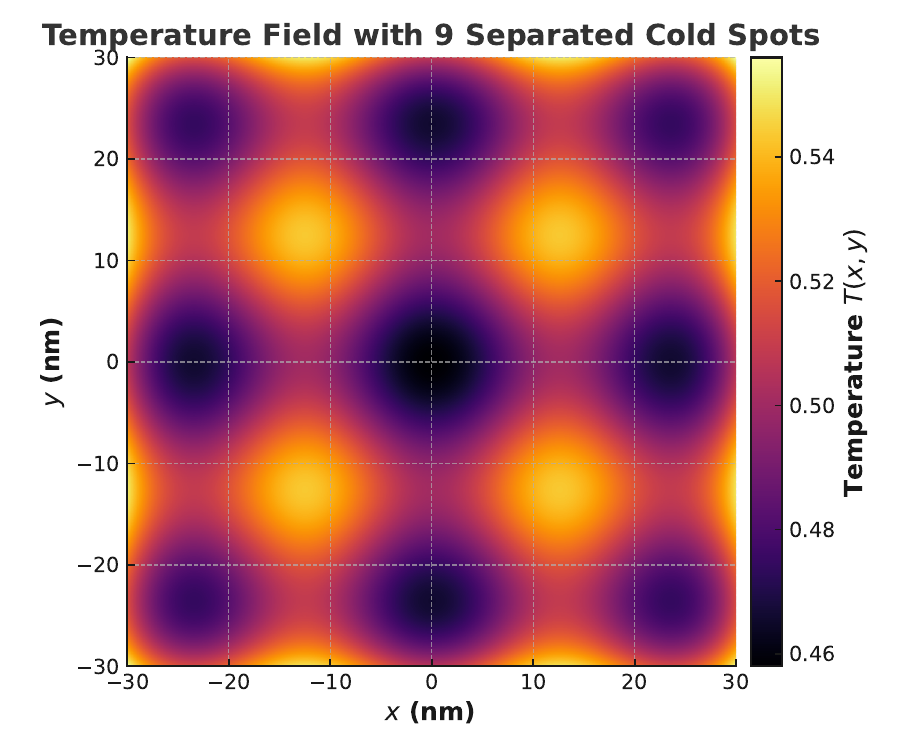}
\caption{
(Color online) Spatial distribution of the temperature field \( T(x, y) \) generated by nine symmetrically arranged cold spots in a 3\(\times\)3 grid. The centers are located at \( (x_i, y_i) = \{ \pm 25, 0 \} \,\mathrm{nm} \), \( (0, \pm 25) \,\mathrm{nm} \), and \( (\pm 25, \pm 25) \,\mathrm{nm} \), with a spatial width \( \sigma = 10 \,\mathrm{nm} \). The field follows \( T(x, y) = T_0 / [1 + \sum_i \exp( -((x - x_i)^2 + (y - y_i)^2)/(2\sigma^2))] \), where \( T_0 \) denotes background temperature. Such thermal modulation can be experimentally realized using thermoplasmonic substrates or spatial light modulators. The energy scale \( \Phi \) is assumed to be of the order of \( 0.025\,\mathrm{eV} \), comparable to \( k_B T \) at room temperature.
}
\label{fig:temperature_density}
\end{figure}
Figure~\ref{fig:temperature_density} depicts  the spatial profile of the temperature field \( T(x, y) \) arising from nine symmetrically distributed cold spots arranged in a 3\(\times\)3 square grid. Each cold spot creates a localized temperature depression modeled by a Gaussian of width \( \sigma = 10 \,\mathrm{nm} \), centered at positions \( (x_i, y_i) \in \{ \pm 25, 0 \} \,\mathrm{nm} \). This superposition results in a structured thermal landscape governed by the expression \( T(x, y) = T_0 / [1 + \sum_i \exp( -((x - x_i)^2 + (y - y_i)^2)/(2\sigma^2))] \), where \( T_0 \) is the uniform background temperature. This configuration provides a controlled way to break the thermal symmetry and impose spatial modulation in the diffusive dynamics of impurity particles. As discussed previously, in experimental realizations, such temperature fields can be implemented using thermoplasmonic heating, laser patterning, or spatial light modulation techniques, allowing for direct control over localization and transport in engineered thermal environments.

\begin{figure}[ht]
\centering
\includegraphics[width=0.7\linewidth]{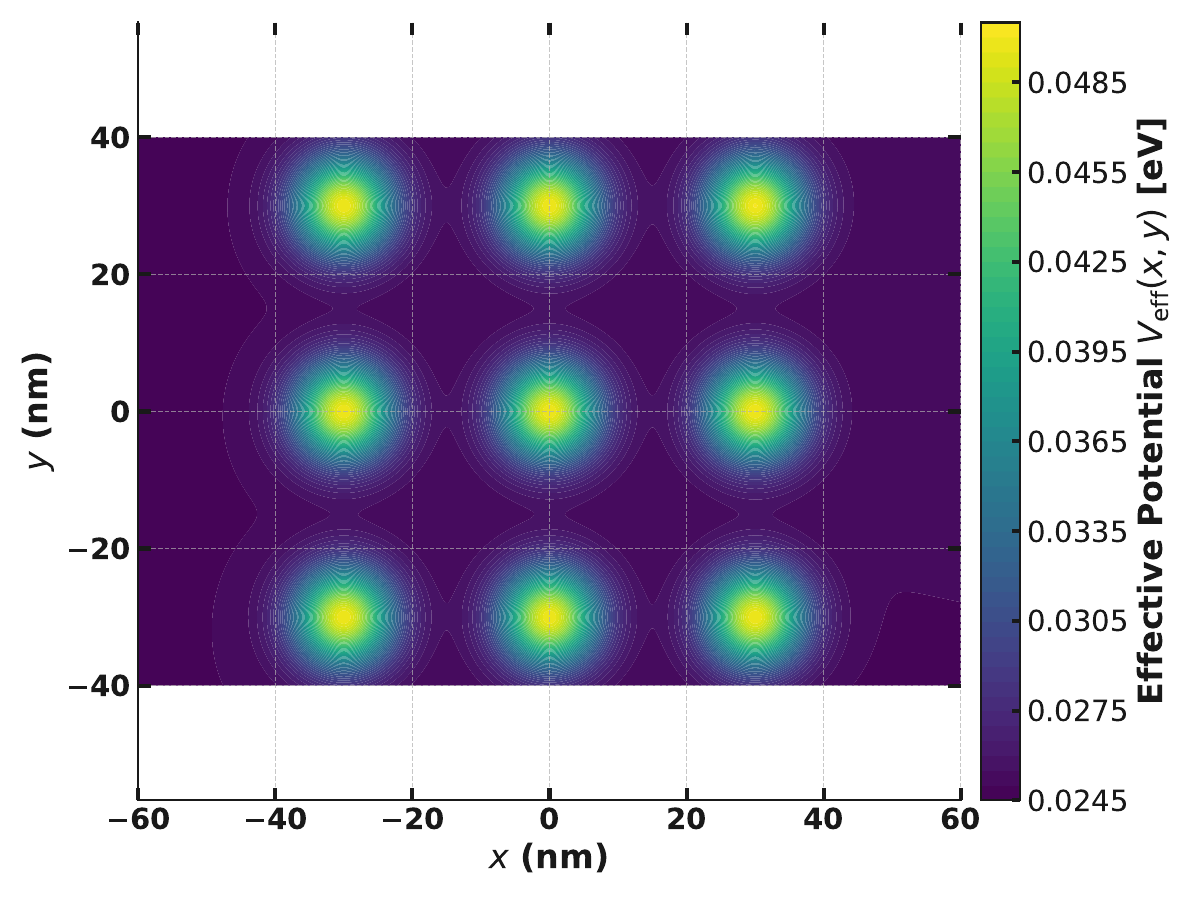}
\caption{
(Color online) Contour plot of the emergent effective potential \( V_{\mathrm{eff}}(x,y) \), defined as the sum of the entropic contribution proportional to \( \Phi T_0 / T(x,y) \) and the spatial integral of the confining potential gradient weighted by the inverse temperature. The temperature field \( T(x,y) \) features nine Gaussian-shaped cold spots arranged in a symmetric 3×3 grid centered at coordinates \( (x_i,y_i) \in \{-30,0,30\} \) nm, each with a spatial width \( \sigma = 5 \) nm. The background temperature \( T_0 \) is normalized to unity, whereas \( \Phi = 0.025\,\mathrm{eV} \) sets the scale of the thermal sensitivity. External confinement is modeled by a quartic potential with amplitude \( V_0 = 0.01\,\mathrm{eV} \) and characteristic lengths \( x_{\max} = y_{\max} = 100\,\mathrm{nm} \). This multi-well potential landscape governs impurity localization and thermally activated transitions in an inhomogeneous thermal environment.
}
\label{fig:veff_9_cold_spots}
\end{figure}

Figure~\ref{fig:veff_9_cold_spots} depicts the emergent effective potential \( V_{\mathrm{eff}}(x,y) \) that governs the steady-state distribution of impurities that move in a thermally inhomogeneous environment under external confinement. This potential comprises an entropic component inversely related to the spatial temperature field \( T(x,y) \) and an energetic component arising from the quartic confining potential \( V(x,y) \) weighted by local temperature variations. The temperature profile includes nine Gaussian-shaped cold spots arranged symmetrically in a \( 3 \times 3 \) lattice, each characterized by a spatial width of 5 nm and positioned at \((x_i,y_i) \in \{-30, 0, 30\}\,\mathrm{nm}\). The background temperature \( T_0 \) is normalized to unity, whereas the energy parameters \( \Phi \) and \( V_0 \) are chosen in the order of room temperature thermal energy (\(\sim 25\,\mathrm{meV}\)), capturing the relevant physical scale for impurity thermodynamics.

The introduction of multiple independently tunable cold spots within anisotropic quartic confinement substantially extends the capabilities of conventional uniform or single-spot cooling schemes. By tailoring the number, positions, and widths of these spots, one can engineer a highly reconfigurable thermal landscape in which the local entropic bias and global confinement act synergistically to direct impurity motion. This versatility enables rapid switching between complex potential topographies, ranging from isolated double wells to extended periodic arrays, without altering the underlying material structure. In such landscapes, thermal noise can be harnessed to activate stochastic‐resonance–mediated transport, thereby boosting dopant mobility and amplifying weak external signals. The resulting noise-assisted dynamics provide a practical pathway for precise doping control, effectively translating the theoretical predictions of noise-enhanced transport into experimentally accessible semiconductor and optothermal platforms.

{\it Results and Summary. \textemdash} The interplay between spatial thermal gradients and external confinement fundamentally governs impurity localization and transport in semiconductor structures. Our results validate and extend those of previous studies~\cite{18,19,32}. Our study   shows that when the cold-spot width \(\sigma\) increases, it  gradually transforms the effective potential \(V_{\mathrm{eff}}(x,y)\) from a bistable landscape (characterized by two peripheral wells and a central barrier) into a monostable configuration. For narrow cold spots (\(\sigma < x_0\)), the central barrier facilitates noise-induced hopping and supports stochastic resonance ~\cite{1,2,3,4}. As \(\sigma\) approaches or exceeds \(x_0\), these wells coalesce, and bistability is lost due to the flattening of the intervening barrier.

The strength of confinement, quantified by \(\Phi\), plays a complementary role in modulating the depth and sharpness of wells. Larger \(\Phi\) values enhance impurity localization near cold regions and increase the activation barriers for hopping. Thus,  \(\Phi\)   independently controls both equilibrium distributions and nonequilibrium transport dynamics. This tunability makes the system highly suitable for optimizing thermally driven ratchet efficiency and amplifying weak signals  \cite{21,25}.

Notably, the introduction of two symmetrically arranged cold spots helps to manuplate  dopant diffusion  across the device. By tuning the separation  between the two cold sopts, and  \(\sigma\), one can move impurities at desired locations and  simultaneously  modulate the potential landscape to enable or suppress transport pathways. This dual-temperature configuration  exposed  to anisotropic quartic confinement creates spatial energy funnels and entropic barriers that guide dopants with minimal external force input. Compared with uniform or single-spot cooling strategies, this approach enables the dynamic reconfigurability of the potential landscape and significantly enhances the range of accessible doping profiles.

These phenomena can be readily implemented using existing experimental tools. External trapping can be realized via gate-defined electrostatic potentials or optically projected quartic traps, whereas the thermal landscape can be sculpted using focused laser heating, spatial light modulators, or thermoplasmonic substrates~\cite{33,34,35,36}. Such methods allow submicron spatial control of both temperature and potential, facilitating experimental access to noise-activated transport and bistability-to-monostability transitions.

To probe impurity distributions, high-resolution techniques such as secondary ion mass spectrometry (SIMS), scanning electron microscopy coupled with energy-dispersive X-ray spectroscopy (SEM–EDX), and spatially resolved photoluminescence (PL) mapping may be employed. In addition, stochastic resonance can be detected through lock-in amplification or time-resolved PL measurements under weak periodic forcing~\cite{32}. Together, these findings establish a broadly applicable framework for manipulating impurity dynamics in semiconductors through joint thermal and mechanical landscape design, offering new pathways for dopant control, noise-enhanced sensing, and energy-efficient information processing.

\section*{Acknowledgment}
MT would like to thank  Mulu  Zebene and Asfaw Taye for the
constant encouragement. 

\section*{Data Availability Statement }This manuscript has no
associated data or the data will not be deposited. [Authors’
comment: Since we presented an analytical work, we did not
collect any data from simulations or experimental observations.]

\section*{Author Contribution Statement }
All authors contributed equally to the conception, 
development, analysis, and writing of this work.

\section*{Appendix A: Derivation of the Effective Activation Barrier and Scaling Laws}

We consider a two-dimensional system of non-interacting impurities diffusing in a spatially inhomogeneous thermal field and subjected to an external trapping potential. The temperature profile comprises two symmetric Gaussian cold spots located at \( (x, y) = (\pm x_0, 0) \), modeled as
\begin{equation}
T(x,y) = \frac{T_0}{1 + \exp\left[-\frac{(x - x_0)^2 + y^2}{2 \sigma^2}\right] + \exp\left[-\frac{(x + x_0)^2 + y^2}{2 \sigma^2}\right]},
\end{equation}
where \( T_0 \) denotes the far-field background temperature, \( \sigma \) characterizes the cold-spot width, and \( x_0 \) defines the lateral position of the spots.

The maximum fractional temperature drop is defined by
\begin{equation}
\epsilon \equiv \frac{T_0 - T_{\min}}{T_0} \in [0,1],
\end{equation}
where \( T_{\min} = T(x_0, 0) \) is the local temperature minimum at the cold spot center. In the limit of well-separated cold spots (\( x_0 \gg \sigma \)), we approximate
\begin{equation}
T_{\min} \approx \frac{T_0}{2}, \quad \Rightarrow \quad \epsilon \approx \frac{1}{2}.
\end{equation}

The steady-state impurity distribution obeys
\begin{equation}
P_{\mathrm{ss}}(x,y) = \frac{1}{Z} \exp\left(-\frac{V_{\mathrm{eff}}(x,y)}{k_B T_0}\right),
\end{equation}
where the effective potential is expressed as
\begin{equation}
V_{\mathrm{eff}}(x,y) = \frac{\Phi T_0}{T(x,y)} + k_B T_0 \int_{(x_0,0)}^{(x,y)} \frac{\nabla V(x',y')}{T(x',y')} \cdot d\mathbf{r}'.
\end{equation}

Here, \( \Phi \) denotes the amplitude of the trap-induced correction and \( V(x,y) \) is the confining potential.

Expanding the temperature near the background as \( T(x,y) \approx T_0 \left(1 - \epsilon f(x,y)\right) \), with
\begin{equation}
f(x,y) = \exp\left[-\frac{(x - x_0)^2 + y^2}{2 \sigma^2}\right] + \exp\left[-\frac{(x + x_0)^2 + y^2}{2 \sigma^2}\right],
\end{equation}
the inverse temperature admits a first-order approximation
\begin{equation}
\frac{1}{T(x,y)} \approx \frac{1}{T_0} \left(1 + \epsilon f(x,y)\right) + \mathcal{O}(\epsilon^2).
\end{equation}

Evaluating the trap-induced energy difference between the cold spot center and the origin yields
\begin{equation}
\Delta V_{\mathrm{trap}} = \Phi \left( \frac{1}{T(0,0)} - \frac{1}{T(x_0, 0)} \right) \approx \Phi \epsilon \left[ 2 e^{-x_0^2/(2\sigma^2)} - 1 \right].
\end{equation}

Assuming a quartic confining potential
\begin{equation}
V(x,y) = V_0 \left[ \left(\frac{x}{x_{\max}}\right)^4 + \left(\frac{y}{y_{\max}}\right)^4 \right],
\end{equation}
the corresponding gradient integral contribution to the barrier can be approximated as
\begin{equation}
\Delta V_{\mathrm{conf}} = k_B T_0 \int_{(x_0,0)}^{(0,0)} \frac{\nabla V(x',y')}{T(x',y')} \cdot d\mathbf{r}' \approx -\frac{V_0 k_B x_0^4}{x_{\max}^4},
\end{equation}
where the integral path is taken along the \( x \)-axis at \( y=0 \), and temperature variations in the denominator are approximated by \( T_0 \).

Combining both contributions, the effective activation barrier  can be written as  
\begin{equation}
\Delta V_{\mathrm{eff}} = \Phi \epsilon \left[ 2 e^{-x_0^2/(2\sigma^2)} - 1 \right] + \frac{V_0 k_B x_0^4}{x_{\max}^4}.
\end{equation}  
  This equation  exhibits   how the trap depth \( \Phi \), cold spot geometry \((x_0, \sigma)\), and confining potential parameters \( V_0, x_{\max} \)  dictate the activation barrier height.

\end{document}